%% file: main.tex
\documentclass[sigconf]{acmart}
\pdfoutput=1
\usepackage{ascmac}

\newif\ifdraft
\drafttrue
\input{macros}

\def\BibTeX{{\rm B\kern-.05em{\sc i\kern-.025em b}\kern-.08emT\kern-.1667em\lower.7ex\hbox{E}\kern-.125emX}}

\newcommand{\ccgrep}{{\tt ccgrep}}
\newcommand{\grep}{{\tt grep}}
\newcommand{\ccfinderX}{{\tt ccfinderX}}
\newcommand{\tb}{\textbackslash}

\setcopyright{none}
\renewcommand\footnotetextcopyrightpermission[1]{}
\begin{document}

%
\title{Code Clone Matching: \\ A Practical and Effective Approach to Find Code Snippets}
%

\author{Katsuro Inoue}
\affiliation{%
 \institution{Osaka University}
 \city{Osaka}
 \country{Japan}
}
\email{inoue@ist.osaka-u.ac.jp}

\author{Yuya Miyamoto}
\affiliation{%
 \institution{Osaka University}
 \city{Osaka}
 \country{Japan}
 }
 \email{yuy-mymt@ist.osaka-u.ac.jp}

\author{Daniel M. German}
\affiliation{%
 \institution{University of Victoria}
    \city{Victoria}
 \country{Canada}}
\email{dmg@uvic.ca}

\author{Takashi Ishio}
\affiliation{%
 \institution{Nara Institute of Science and Technology}
    \city{Ikoma-City}
 \country{Japan}}
\email{ishio@is.naist.jp}

%
\renewcommand{\shortauthors}{Inoue, et al.}

%
\begin{abstract}

Finding the same or similar code snippets in source code is one of  fundamental activities in software maintenance. Text-based pattern matching tools such as  \grep\ is frequently used for such purpose, but making proper queries for the expected result is not easy. Code clone detectors could be used but their features and result are generally excessive. In this paper, we propose Code Clone matching (CC matching for short) that employs a combination of token-based  clone detection and  meta-patterns enhanced with meta-tokens. The user simply gives a query code snippet possibly with a few meta-tokens and then gets the resulting snippets, forming type 1, 2, or 3 code clone pairs between the query and result. By using a code snippet with meta-tokens as the query, the resulting matches are well controlled by the users.  CC matching has been implemented as a practical and efficient tool named \ccgrep, with \grep-like user interface. The evaluation shows that \ccgrep~ is a very effective to find various kinds of code snippets.

\end{abstract}

%
%


%
%
%
\maketitle

\section{Introduction}

Finding and locating code snippets in source code files are fundamental activities to understand and maintain software systems\cite{Roehm:2012Comprehend,dit2013feature}.
%
When we identify a bug in a code snippet, we would search the same or similar snippets in the same or other systems to check if they contain the same bug\cite{Jiang2007,Nguyen:2010,Jang:2012:ReDeBug}.
When we modify a code snippet for feature change or enhancement, we would find the same or similar snippets for preventing unintentional inconsistent bug\cite{TsantalisRefactorability2015,JDeodorant2016}.
When we identify a code snippet with a bad coding practice, we would search the same or similar snippets for possible refactoring\cite{tairas2012ISTcloneRefactoring}.

To find the same or similar code snippets effectively, various kinds of software engineering tools or IDE's have been proposed and implemented\cite{storey2005comprehention,dit2013feature};
%
however, it has been reported that the character-based pattern matching tool \grep\cite{grepGnu} is still widely used to
find code snippets, due to its simplicity, trustworthiness, speed, and availability\cite{Singer1997, kernighan1999}.
Although \grep~  is very convenient to find lines with specific keyword, it is not easy to make a proper query for a
code snippet that  ignores  comments and  white spaces, and might span multiple lines.

%
%

Code clone is a pair of code snippets those are identical or similar each other\cite{carter1993clone}.
A large body of scientific literature on clone detection has been published and various kinds of code clone detection tools (detectors) have been developed\cite{ROY2009470,rattan2013software}.
Most of these detectors are designed to detect all of the code clone pairs in the target source files, and thus, the
resulting code clone pairs become generally huge\cite{kim2005empirical} and they contain  a lot of code clone pairs that
might not be of interest (including false positives).
There are a few tools that find similar code snippet for a query code snippet, but their performance and usability are limited\cite{LiCBCD2012,IshioMSI18,Chaiyong2019Siamese}.

In this paper we propose a method, which we call \emph{Code Clone matching} (CC matching), to find clones of specific code snippets by using a combination of clone detection and pattern matching.
Search queries can be simply code snippets, or code snippets enhanced with meta-patterns (with meta-tokens having leading \$) that can provide flexibility to narrow or broaden the search query.
An example of a query that uses meta-patterns would be searching for for-loops in
which the control variable of the for-loop is the variable \texttt{index} and which contains a \texttt{return} statement inside the body of this for-loop, represented like

   \texttt{for(\$index=\$\$)\{\$\$ return \$\$\}} \\
As a consequence, CC matching is programming language aware, and able to properly tokenize the source code (ignoring whitespace and comments). In clone-detection terms, CC matching retrieves, given a code snippet (that potentially includes some meta-patterns), the clones (type 1, 2 and 3) of this snippet that also satisfy the meta-patterns (e.g. that use a  specific variable name in some specific locations).


We will also present {\em ccgrep}, that is  current implementation of CC matching for C, C++, Java, and Python. \ccgrep~ works as a handy but reliable pattern matching tool with a grep-like and easy-to-understand user interface.
An evaluation of \ccgrep~ shows that it is  capable of representing
the queries for all of type 1, 2, and a part of type 3 clones. In addition, \ccgrep\
accurately performs CC matching, and the speed is slower than \grep~ but faster than other similar code finders.

As far as we know, little has been studied on the pattern matching based on the notion of  code clone.
The contributions of this paper are twofold:

\begin{itemize}
\item  We propose CC matching, a new concept of matching code snippets based on token-based code clone detection and enhanced pattern matching.

    \item   We present a practical, efficient, and easy-to-use tool called \ccgrep\ that implements CC matching, with its evaluation.
\end{itemize}


\section{Background}

\subsection{Motivating Example}

Some uses of the ternary operator (e.g., \mycode{exp1 ? exp2 : exp3} meaning the result of this entire expression is \emph{exp2} if
\emph{exp1} is true, otherwise the result is \emph{exp3}---available in C, C++ and Java) are, arguably, considered bad
practice\cite{Ternary}. For example, the use of \emph{a < b ? a : b} is arguably harder to read than using
\emph{min(a,b)}. Therefore, it might be desirable to replace the ?: operator with a function or macro that returns the
minimum value.  The following is an example found in the file \mycode{drivers/usb/misc/adutux.c} in the Linux kernel
(v5.2.0).

\begin{description}
\item {\tt  amount = bytes\_to\_read < data\_in\_secondary ?} \\
    \vspace{-0.4cm}
     \flushright{
      {\tt bytes\_to\_read : data\_in\_secondary;}}
\end{description}

\noindent
This  line of code should be replaced with a more readable expression (note that the macro \mycode{min} in Linux
guarantees no side effects):

\begin{description}
\item {\tt amount = min(bytes\_to\_read, data\_in\_secondary);}
\end{description}

Finding all occurrences of such usage of the ternary operator using \mycode{grep} is not easy.
For example, simply executing "{\tt grep '<'}" for all 598 files (total 51,6394 lines in C) under /drivers/usb produces 16335 matching, including many unwilling patterns such as "\mycode{if (a$<$b)}", "\mycode{for (i=0; i$<$x; ...)}", or "\mycode{\#include $<$linux/...$>$}". We could narrow the matches by concatenating \grep~ like,
\begin{description}
\item {\tt grep '<' -r . | grep '?' | grep ':'}
\end{description}
However, it still produces 149 matches. Perhaps more problematic is that the expressions could span multiple
lines. While it is possible to create a complex regular expression to find these expressions, it would be time consuming and potentially error prone.

Another alternative would be to use a clone detector to detect uses of the ?: operator. The  clone detector
NiCad\cite{Cordy2011NiCad} detects 646 block-level clone classes for the \mycode{drivers/usb} files by the default setting, but no snippet with
the ternary operator case is included in the result because it is too small to be detectable.

Ideally we would like to be able to specify simple and easy-to-create-and-understand query to find these types of
snippets. Therefore in this paper we propose CC matching, which is based on the notion of code clone detection. Using CC matching, the query is written simply as:
\begin{description}
\item{\tt a < b ? a : b}
\end{description}
In a nutshell, this query specifies that a variable (represented by a) should be followed by \mycode{<} and then a
second variable (represented by b),
followed by a \mycode{?}, followed by the same first variable found, followed by \mycode{:}, followed by the second
variable. Also, whitespace and comments should be ignored. This query would match \mycode{x<y?x:y} but it would not
match \mycode {x<y?x:z}.

Using this query, we used our implementation of CC matching (which we call \ccgrep ) to find the occurrences of this type Linux's
\mycode{drivers/usb}. We identified 3 instances, and submitted patches to replace them with \mycode{min}. Two of those
patches have been accepted already into Linux.


\subsection{Pattern Matching with grep}

\grep~ takes a query pattern in a form of a regular expression (or an extended regular expression), and reports the matched lines in the target files. Developers  specify a keyword, or a short idiom as the query, and get the result output composed of the file paths and matched lines.
\grep\ is generally fast and effective, but sometimes a simple query pattern  generates a large output which is hard to
analyze further. A complex query pattern might reduce the size of the output, but making a proper query is not an easy
task even for an expert of the tool.

Also, \grep\  is designed to work with any text file, and it is not specifically designed to explore source code
files. Unless a complex query is written, it reports matches in comments (which users might want to have ignored) and
it is difficult to find matches that spawn multiple lines. Furthermore, because grep is based on finite state machines,
it is not capable of dealing with matching parenthesis, brackets or braces (e.g. one might want to match the entire
block of code inside a for-loop, including other blocks inside it).
There have been various proposals for extending \grep~ to do code matching\cite{Dekel2003-pattern-lang, sgrep2002,
  AbouAssaleh2004SurveyOG}; however, none of these have been successful as \grep.





\subsection{Code Clone Detection and Search}
\label{sec:detector}
A code clone is broadly defined as a code snippet having the same or similar code snippets in the target software
collection\cite{ROY2009470,Kamiya2002CCFinder,Baker92aprogram}, and a code clone pair is classified into 4 types, type 1
(syntactically the same snippets except for comment or whitespace differences), type 2 (type 1 with identifier or literal
differences), type 3 (type 2 with addition, deletion, or change of statements within the clone), and type 4 (semantically equivalent
snippets)\cite{ROY2009470}.


Code clone detector is a promising method for finding code snippets with unique characteristics. A user would want to simply
provide a snippet and then find all the clones of this specific snippet.  However, most of those clone detectors report
all of code clone pairs in the target files; these results are generally huge and mostly irrelevant for finding specific
code snippets. Further, most clone detector tend to ignore small-size clones\cite{mondai2018microclones} and so they would miss small code snippet for search as shown in the motivating example.

Some code clone detector, such as \ccfinderX~\cite{ccfinderx}, have an option to find clones between a specified file and all other files; however,
we still need to prepare a query file and also need to tune various parameters to the specifics of the query (such as
its length), which heavily affect the detection result\cite{ROY2009470}.

There are several tools dealing with code search for code clone pairs, such as CBCD\cite{LiCBCD2012}, NCDSearch\cite{IshioMSI18}, and Siamese\cite{Chaiyong2019Siamese}. CBCD is a PDG (Program Dependency Graph)-based code matching tool by graph isomorphism testing. NCDSearch is a block-based code search tool using normalized compression distance of two code snippets. Siamese is token-based code retrieval tool with inverted index.  These tools would show good accuracy as code clone finder for certain conditions, but they would still have issues on performance or usability as daily-used practical software engineering tool.


Another issue is usability of the tools. Most of those tools are stand-alone
in the sense that they take their specific input format and generate proprietary output. Integration with
other tools needs to transform their input/output formats, so it becomes hard to collaborate with others.


\begin{table*}[tb]
    \centering
    \caption{Token-Level Matching}
    \raggedright \hspace{.3cm}
    \begin{tabular}{|l|l||l|l|} \hline
      token(s) in query     & matched token(s) in target                                 & \multicolumn{2}{|c|}{simple example of match} \\ \cline{3-4}
                            &                                                            & query         & target                        \\ \hline \hline
      reserved word{\dag}   & exact reserved word                                         & \tt{while}    & \tt{while}                    \\ \hline
      delimiter             & exact delimiter                                             & \tt{(}        & \tt{(}                        \\ \hline
      identifier            & any identifier{\ddag}                                      & \tt{myname}   & \tt{abc}                      \\ \hline
      literal               & any literal{\ddag}                                         & \tt{1}        & \tt{100}                      \\ \hline
      \$identifier          & exact identifier                                           & \tt{\$myname} & \tt{myname}                   \\ \hline
      \$literal             & exact literal                                              & \tt{\$1}      & \tt{1}                        \\ \hline
      \$.                   & any single token                                           & \tt{\$.}      & \tt{if}                       \\ \hline
      \$\#\ \ X             & any shortest token sequence ending with X                           & \tt{\$\# +}   & \tt{while(f(a+}                \\ \hline
      \$\$\ \ X             & any shortest token sequence ending with X, excluding X  &              &                               \\
                            & inside well-balanced bracket  \{...\}, [...], or (...)                & \tt{\$\$ +} & \tt{while(f(a+1))+}                   \\ \hline
      X\ \ \$|\ \ Y         & either  X or Y                                             & \tt{+ \$| -}  & \tt{-}                        \\ \hline
      X\ \$*                & repeated sequence of X zero or more times                  & \tt{( \$*}    & \tt{(((}                       \\ \hline
      X\ \$+                & repeated sequence of X one or more times                   & \tt{( \$+}  & \tt{((}                       \\ \hline
      X\ \$?                & X or none                                                 & \tt{( \$?}    & \tt{(}                        \\ \hline
      \$( ~~X1 X2 ... ~~\$) & X1, X2, ...  (group for further regular expression operations)
                            & \tt{\$( a++ \$| ++a \$)}                                   & \tt{a++}                                      \\ \hline
    \end{tabular}                                                                                                                        \\
    \raggedright \hspace{.3cm}
{\dag}Type names are treated as identifiers.\\
\hspace{.3cm} {\ddag}Identifier and literal may match only the exact one by an option.\\
\hspace{.4cm}Tokens starting with \$ are meta-tokens and others are regular tokens.\\
\hspace{.4cm}Wildcard meta-tokens \$\# and \$\$ match in reluctant way, and \$*, \$+, and \$? match in possessive way\cite{JavaRegularExpressions}.\\
\hspace{.4cm}X, Y, X1, X2, ... are any regular token or a group with \$( ... \$).
\label{tab:tokens}
\end{table*}

\section{CC Matching}
\subsection{Design Policy}

Our design policies of CC matching are as follows.

\begin{itemize}

    \item CC matching is a process of finding code snippets in the target source code for a query snippet. The query and
      matched result can be seen as a code clone pair of either type1, 2, or 3. The query  is a code snippet or a code
      snippet enhanced with meta-patterns that can widen or narrow the potential matches.

    \item The matching is made at a granularity of a sequence of language-dependent tokens (white space and comments are
      removed) and not as simply sequences of characters.
. 

    \item CC matching can also precisely control how tokens in the query are matched.  Using command line options, the query pattern can match type 1, type 2 (P-match), type 2 (non-P-match) and a part of type 3 clone snippets effectively.

\end{itemize}

\subsection{Formulating  CC Matching}
\label{sec:BasicFeaturesofCC}

The input of CC matching is the query  $q$, the target $T$ of source code files in a programming language $L$, and
matching option $o$. The output is a matched code snippet $t$ in $T$\footnote{Note that in an actual implementation of CC matching, the location of all the matched code snippets in $T$ will be the output, but for simplicity of the explanation here, we take one of the matched snippets itself $t$ as the output.}.
%
We refer to reserved words, delimiters (operators, brackets, ; ...), identifiers, and literals in $L$ as \emph{regular
  tokens}. Other tokens starting with meta symbol \$ are called \emph{meta-tokens}.
$q$ is a sequence of regular tokens and the meta-tokes, and each matched result $t$ is a sequence of the regular tokens.
These token sequences do not contain comments, white spaces, or line breaks. We always consider the matching on the
token sequence level, not on the character level.


In Tab.\ref{tab:tokens}, we define a token-level matching for various kinds of tokens in CC matching, with simple examples. The basic ideas of these matches are as follows.
\begin{itemize}
    \item A language-defined token such as reserved words or delimiters matches the exact token.
    \item A user-defined token such as identifier or literal can match same kind of token with a possibly different name
      or value. To pin down them to a specific identifier name or literal value, \$ is used before the token. For
      example, \mycode{\$count} would match only the token \mycode{count}.
    \item Wildcard tokens \$., \$\#, and \$\$ are introduced for the matches to any single token, any token sequence,
      or any token sequence discarding paired brackets, respectively.
    \item Popular regular expression operators for choice, repetition, and grouping are introduced to enhance the expressiveness.
\end{itemize}

Consider that query $q$ is a  token sequence $q_1, ... , q_m$ ($1 \leq m$), and a target $t$ is a token sequence $t_1, ... , t_n$ ($0 \leq n$).
From $q_1$ to $q_m$, if each token in the query matches tokens in the target from $t_1$ to $t_n$ as defined in Tab.\ref{tab:tokens} without overlapping or orphan tokens, then we say $q$ matches $t$ by CC matching.



\subsection{Type 1 Matching}

The most simple type of query in CC matching---named {\em type 1 matching}---is the case  that $m=n$ and $q_i = \$t_i$
for each  identifier or literal and $q_i = t_i$ for other tokens. Type 1 matching is performed to find the exact code
snippet, discarding comments, white spaces, or line breaks. This type of query does not use wildcard nor regular
expression tokens. These are
examples of  queries and targets:

\begin{description}
\item[q1:] {\tt \$a = \$0; \$b = \$10;}
\item[t1:] {\tt a = 0; b = 10;}
\end{description}
q1  matches  t1 as type 1 matching. Note that annotating all identifiers and constants with \$ would be sometimes bothersome;
for this reason our implementation has an option  for automatic annotation of these tokens to find any type-1 clone of the query.

The next example is a case of "no match".
\begin{description}
\item[q2:] {\tt \$a = \$0; \$b = \$10;}
\item[t2:] {\tt a = 0; b = 20;} \hspace{0.25cm} (no match)
\end{description}
Literal \texttt{10} does not match \texttt{20}, thus  overall  q2 does not match t2.

\subsection{Type 2 Matching, P-Matching, and Pinning Down}

For the query token sequence $q_1, ..., q_m$ and the target token sequence $t_1, ..., t_n$, if $n=m$ and $norm(q_i) = norm(t_i)$ for each $i$,  then $q$ matches $t$ as \emph{type 2 matching}. Here $norm$ is a normalization function to flat the distinction of identifiers (or literals), defined below.\\

\( norm(x) \equiv \left\{ \begin{array}{ll}
                       \#id  & \mbox{if $x$ is an identifier}\\
                       \#li  & \mbox{if $x$ is an literal}\\
                       x     & \mbox{otherwise}
                       \end{array}
                       \right.
                       \) \\
In type 2 matching, an identifier in the query can match any identifier in the target, and also a literal in the query can match any literal in the target.

\begin{description}
\item[q3:] {\tt a = 0; b = 10;}
\item[t3:] {\tt x = 10; y = 200;}
\end{description}
q3 matches t3, because the sequences of the normalized tokens are both $[\#id, =, \#li, ;, \#id, =, \#li, ;]$.

A special case of type 2 matching, with a constraint such that for any identifier or literal $q_i$ if $q_i = q_j$, then $t_i = t_j$, is called {\em Parameterized match} or \emph{P-matching}. This is sometimes referred to consistent or aligned matching, meaning the same identifiers (or literals) in the query are mapped into the same ones in the target. P-matching is formally defined with a specialized normalization function $norm_p()$, as follows.\\

\(norm_p(x) \equiv \left\{ \begin{array}{ll}
                       \#id_{pos(x)}  & \mbox{if $x$ is an identifier}\\
                       \#li_{pos(x)}  & \mbox{if $x$ is a literal}\\
                       x     & \mbox{otherwise}
                       \end{array}
                       \right.
                       \) \\
Here, $pos(x)$ is a function returning position $i$ such that identifier (or literal) $x$ is the $i$-th identifier (literal) newly appeared in the token sequence\footnote{Note that any meta-token starting with \$ in the query and their matched tokens in the target are out of consideration of $pos()$.}.

\begin{description}
\item[q4:] {\tt a = 0; a = a + b;}
\item[t4:] {\tt y = 0; y = y + c;}
\end{description}
For q4, $pos(a)=1$ and $pos(b)=2$, and for t4, $pos(y)=1$ and $pos(c)=2$.
q4 matches t4 as P-matching, because the P-normalized sequences are both $[\#id_1, =, \#li_1, ;, \#id_1, =, \#id_1, +, \#id_2, ;]$. The following case is type 2 matching but not P-matching.
\begin{description}
\item[q5:] {\tt a = 0; a = a + b;}
\item[t5:] {\tt y = 0; y = z + c;}  (type 2 matching but not P-matching)
\end{description}
At t5, \texttt{z} cannot be matched by \texttt{a} because $norm_p(a) = \#id_1$ is not equal to $norm_p(z)=\#id_2$. As a default of CC matching, P-matching is assumed\footnote{It can be changed by the tool's option}.

Sometime type 2 matching, even P-matching might match many targets, and we would  want to narrow them by pinning down an identifier (or literal) to a specific one. To do this, we also use the annotation of \$ at the beginning of the identifier (literal) in the query, so that the identifier (literal) headed by \$ is not normalized and it matches only the exact token in the target.
\begin{description}
\item[q6:] \texttt{\$cat = 1}
\item[t6-1:] \texttt{cat = 1} (match)
\item[t6-2:] \texttt{dog = 1} (no match)
\end{description}
q6 matches t6-1, but it does not match t6-2 since identifier \texttt{cat} is not normalized and  is pin down to \texttt{cat}.

We can mix normalized identifiers (literals) and non-normalized ones in the query as follows.
\begin{description}
\item[q7:] \texttt{\$cat = cat+1}
\item[t7:] \texttt{cat = dog+1}
\end{description}
q7 matches t7, because \texttt{\$cat} in q7 and its corresponding target \texttt{cat} in t7 are excluded from consideration of $pos()$, and so both \texttt{cat} in q7 and \texttt{dog} in t7 are treated as the first appeared identifiers and both are normalized to $\#id_1$.

\subsection{Wildcard Tokens and Type 3 Matching}

There are three  wildcard tokens, \$. (any token, but only one), \$\# (any token sequence), and \$\$ (any token
sequence with bracket discarding).   \$. matches any single token in the target, and \$\#~X matches any shortest token sequence ending with X.  \$\$~X is similar to \$\#~X but it does not end at X inside balanced brackets \{...\}, [...], and (...). In this case, the match continues until the first X outside the balanced brackets; thus X's inside the brackets are not seen.


For the query token sequence $q_1, ..., q_m$ and the target token sequence $t_1, ..., t_n$, if a wildcard token $q_i$ in the query matches some target tokens $t_j, ..., t_{j+k-1}$, and other query tokens match properly the target tokens (as defined in Tab.\ref{tab:tokens}), preserving the order of the query and the matched target tokens without any overlapping or orphan tokens, we call it \emph{type 3 matching}. Here, $k \geq 0$, meaning that the matched token sequence may be none($k=0$), a single token($k=1$), or more($k \geq 2$). Here we present several examples of wildcard tokens.

\begin{description}
\item[q8:] \texttt{\$a = \$. ;}
\item[t8-1:] \texttt{a = b ;}
\item[t8-2:] \texttt{a = 10 ;}
\end{description}
In this case, query q8 matches both t8-1 and t8-2, by the replacement of \$. with \texttt{b} and \texttt{10}, respectively.

\begin{description}
\item[q9:] \texttt{\$a \$\# ;}
\item[t9-1:] \texttt{a = b+c ;}
\item[t9-2:] \texttt{a++ ;}
\end{description}
Query q9 matches \texttt{a} and any tokens before `\texttt{;}', and \$\# matched `\texttt{= b+c}' in t9-1 and `\texttt{++}' in t9-2.

\begin{description}
\item[q10:] \texttt{\$a = \$\$ \$b}
\item[t10-1:] \texttt{a = b}
\item[t10-2:] \texttt{a = 10+c+b}
\item[t10-3:] \texttt{a = f(b,10)+b}
\end{description}
t10-1 is the case that the wildcard \$\$ matches none, t10-2 is `\texttt{10+c+}', and t10-3 is `\texttt{f(b,10)+}' where the first \texttt{b} is inside the bracket (...) and it is not matched by \texttt{\$b}. The next is a more complex example.

\begin{description}
\item[q11:] \texttt{a= f(p); if(\$\$)\{a=-a;\}}
\item[t11-1:] \texttt{b= g(q); if(c<=0)\{b=-b;\}}
\item[t11-2:] \texttt{b= g(q); if(h(c)==0)\{b=-b;\}}
\end{description}
In this query, the conditional expression of {\tt if} statement can be any token sequence followed by the closing bracket {\tt )}. In t11-1, \$\$ matches {\tt b<=0}, and in t11-2 it matches {\tt h(b)==0} where the first closing bracket is balanced with the open bracket so that the matching for \$\$ continues after {\tt h(b)} until another closing bracket appears.

With these wildcard tokens, we can specify  possible changes of the target token sequence, i.e., variants of the  target code snippet.

\subsection{Regular Expression Extensions }

To effectively represent queries for largely different variants, CC matching employs regular expressions in its query with meta symbols |, *, +, ?, (, and ) preceded by \$.

\begin{description}
    \item[Selection:]
    $p1$ \$| $p2$ means a matching by either patterns $p1$ or $p2$.
    \item[Iteration:]
    $p$\$*, $p$\$+, and $p$\$? are the repetition of $p$ zero or more times, one or more times, and zero or one time, respectively.
    \item[Grouping:] \$( $p$ \$) defines the scope and precedence of meta symbols in pattern. For example,
    \$($p1$ | $p2$ \$)\$+ means any repeated pattern of $p1$ or $p2$.
\end{description}
For example, we can find nested if-else clauses as follows.
\begin{description}
\item[q12:] \texttt{\$( if \$\$ else \$) \$+}
\item[t12:] \texttt{ if(i<10) \{a=0;\} else if(i<20) \{a=5;\} else}
\end{description}
In this case, conditional expressions and then-clauses are matched by \$\$, and if-else are sought until else is not followed by if.


\subsection{Finding Various Code Patterns}

Combining the regular tokens and meta-tokens in the query, we can find various kinds of code snippets in the target, from simple to complex code patterns. The following are examples in Java.

\begin{description}
    \item[- Method $XYZ$ with no parameter]\mbox{}\\
    {\bf q13:} {\tt \$XYZ( )}\\

    \item[- Method $XYZ$ with 0 or more
    parameters]\mbox{}\\
    {\bf q14:} {\tt \$XYZ(\$\$)}\\

    \item[- Method $print$ with variable $buf$ as the 1st parameter]\mbox{}\\
    {\bf q15:} {\tt \$print(\$buf, \$\$)}\\

    \item[- Any method definition]\mbox{}\\
    {\bf q16:} {\tt T f(\$\$)\{\$\$\}}\mbox{}\\
    Note that type names are treated like identifiers and $T$ matches any type name.\\

    \item[- Getter method]\mbox{}\\
    {\bf q17:} {\tt T f()\{return this.v;\}}\\

    \item[- Setter method]\mbox{}\\
    {\bf q18:} {\tt T1 f(T2 v1)\{this.v1=v2;\}}\\

    \item[- $if$ statement]\mbox{}\\
    {\bf q19:} {\tt if (\$\$)\{\$\$\}}\\

  \item[- $for$ statement using control variable]\mbox{}\\
    {\bf q20:} {\tt for(T i=0; i<\$\$; i++)\{\$\$\}}\mbox{}\\
\end{description}

These queries can be narrowed by restricting to specific identifiers or constants, by using \$id, or \$0 to match
\mycode{id} or \mycode{0} exactly.



\section{Implementation of CC Matching}

\begin{figure}
    \centering
    \includegraphics[width=9.0cm]{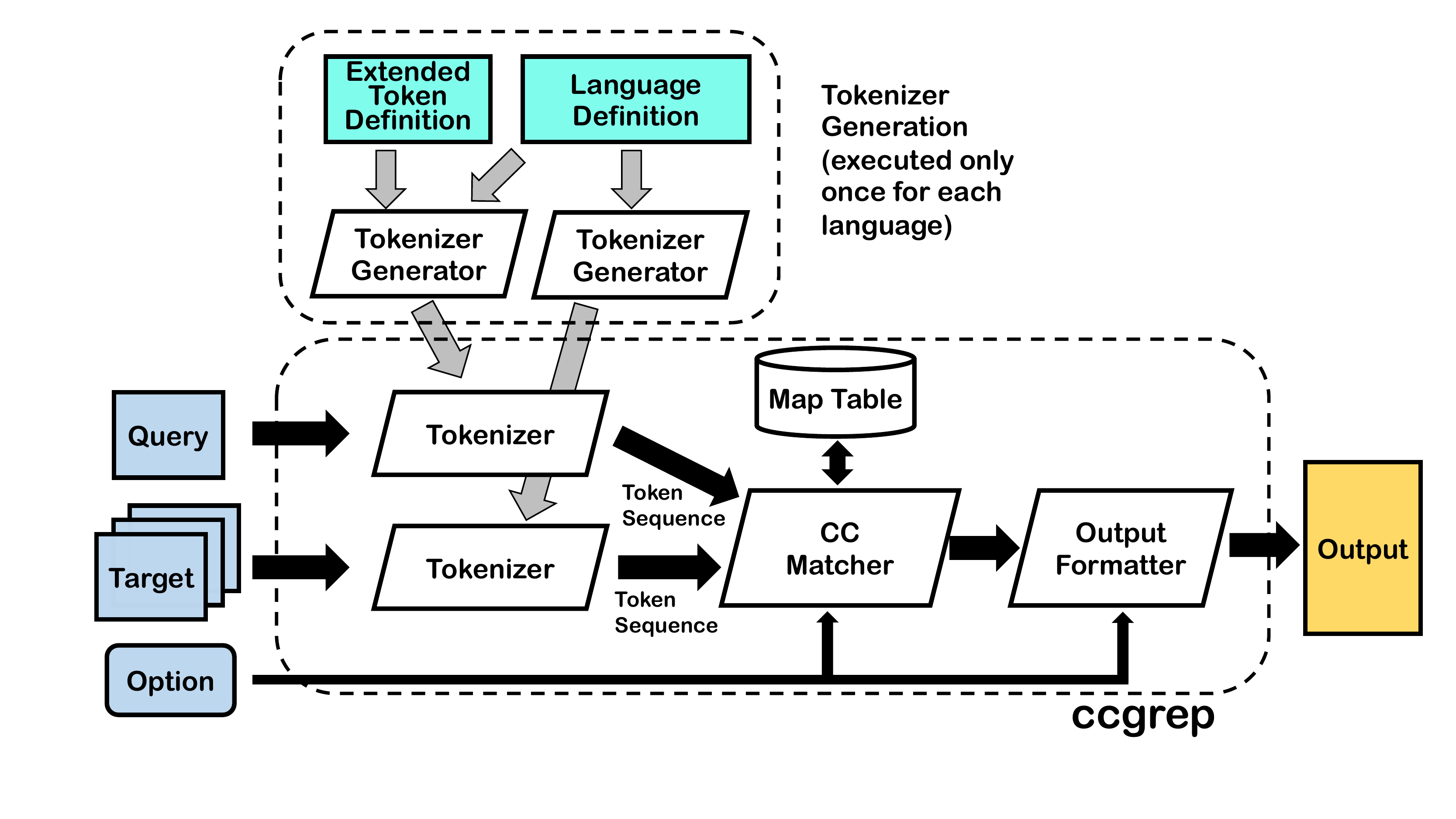}
    \vspace{-1.0cm}
    \caption{Architecture of ccgrep}
    \label{fig:overview}
\end{figure}

We have implemented our proposed CC matching  approach for finding code snippets
in a
tool named {\tt ccgrep}\footnote{https://github.com/yuy-m/CCGrep}. The target languages of \ccgrep~ at this moment are C, C++, Java, and Python3.
We have chosen  a \grep-like input/output interface to facilitate its adoption.

\subsection{Architecture of ccgrep}

The architecture of the process of {\tt ccgrep} is presented in Fig.\ref{fig:overview}. Each component is described below.

\begin{description}

    \item[Tokenizer Generators:] Parser generator ANTLR\footnote{\url{https://www.antlr.org/}} is used to generate two kinds of tokenizers. For the target tokenization, only the language definition is  used to recognize the regular tokens, but for the query tokenization, the definition of the meta symbol extension for the meta-tokens explained in Sec.\ref{sec:BasicFeaturesofCC} and that of regular tokens are used. 
    This process has been executed only once for each target language.

    \item[Tokenizers:] Each tokenizer removes white spaces and comments from the input text, and decomposes the code into tokens. The query tokenizer accepts the meta-tokens starting with \$ and the regular tokens defined by the language, but the target tokenizer accepts only the regular tokens. The tokenizer for the target  files are executed in parallel for each file, along with following CC Matcher.

    \item[CC Matcher:] This performs a naive sequential pattern matching algorithm between two token sequences for the query  and the target\cite{Gusfield97}. For the case of the selection pattern in the query, simply each case is tested one by one.  %
    For type 2 code clone matching, we record $norm_p()$ values for each identifier and literal in Map Table so that we can check if correspondence of identifiers in the query and target is consistent or not. Note that the table contents are flushed for each query.
    The  option  controls the  normalization level,  input language, output form, and many others discussed in the next section.

    \item[Output Formatter:] This process constructs the output for the successful matching result. Based on the input option, we can view the match result, like {\tt grep}, in the form of the file name associated with the matched top line as the default, or as many other styles such as full matched lines, only the number of lines, or so on. Fig.\ref{fig:output} is an example output of {\tt ccgrep} where {\tt catch} statement followed by identifier {\tt IOException} and a specific call {\tt toolError(...)} are sought in java files of ANTLR Ver.4, and the top lines of the matched results with their file paths are listed.

\end{description}

\begin{figure}
\begin{center}
\begin{screen}
\footnotesize
\begin{verbatim}
$ccgrep 'catch($IOException $$){$$ $toolError($$);}' -r .
./parse/TokenVocabParser.java:			catch (IOException ioe) {
./Tool.java:                catch (IOException ioe) {
./Tool.java:		catch (IOException ioe) {
./Tool.java:                catch (IOException ioe) {
./Tool.java:		catch (IOException ioe) {
./codegen/CodeGenerator.java:		catch (IOException ioe) {
./codegen/target/SwiftTarget.java:        catch (IOException ioe) {
$
\end{verbatim}
\end{screen}
\end{center}
  \rightline{The target is ANTLR V.4, {\tt {\footnotesize $\sim$antlr4/tool/src/org/antlr/v4/.}}}
  \vspace{-0.3cm}
\normalsize
    \caption{An Example of {\tt ccgrep} Output}
    \label{fig:output}
\end{figure}

{\tt ccgrep} is written in Java associated with the ANTLR output, and it is easily installed and executed in various
environments such as Unix and Windows (a single JAR is provided that contains all the required libraries needed to run it).

\subsection{Input/Output and Options of ccgrep}

{\tt ccgrep} has a character-based  user interface, and it takes similar options and  standard input/output treatment to \grep\ so that developers familiar with {\tt grep} can easily use {\tt ccgrep}. The options are as follows.
%

\begin{description}

    \item[Target Language:] Currently, Java, C, C++, and Python3 are the target languages we have implemented. The default target language is Java.
    \item[Target Files:] The target files are specified by the command line or they are sought recursively from the specified directory. By connecting \ccgrep\ with pipe '|', the target becomes the result of the previous command, and therefore we can combine \ccgrep\ with other shell commands or \ccgrep\ repeatedly effectively.
    \item[Query Pattern:] The default setting requires the query pattern at the command line. This can be changed to a specified file or the standard input.
    \item[Normalization (Blind) Level:] We can choose the normalization level (blind level) of user defined identifiers and literals, such as none for non normalization of type 1 clone detection, P-matching for type 2 with consistent name change, non-P-matching for all type 2 clone. The default is P-matching.
    \item[Output (Print-Out) Form:] Various kinds of output forms can be chosen. The default is a familiar form of \grep , and this can be changed to different styles. The print out is usually made to the character-based standard output, but it can be changed to a JSON or XML format file.

\end{description}

\section{Evaluation}

Goal of the evaluation is to show that our proposed approach, CC matching and its implementation \ccgrep, can find various kinds of intended code snippets effectively and efficiently, compared to other approaches.  This goal could be decomposed into following three research questions.

\begin{description}

\item[RQ1:Query Expressiveness] Are various kinds of queries expressed by CC matching (and also its implementation \ccgrep )? Also, Are the queries written more easily than  \grep\ ?

\item[RQ2:Accuracy of ccgrep] Is \ccgrep\ accurately find various types of code clones already detected by other approaches?.

\item[RQ3:Performance of ccgrep] What it the execution time of \ccgrep? Is the token-based naive sequential pattern matching approach  fast enough in practice, compared to other tools such as \grep\ or code-clone search tool NCDSearch?

\end{description}

\subsection{RQ1: Query Expressiveness and Effectiveness}

RQ1 explores how easily we can make the queries by our approach.
%
Here we discuss the way of creating queries of each matching type. Since expressiveness of CC matching is equivalent of that of \ccgrep\, we only mention  \ccgrep\ here.

\subsubsection{Various Queries Classified with Matching Types}

As shown in previous sections, it is obvious that our approach can easily create various query patterns for type 1 matching, type 2 matching with P-match, and type 2 matching with non-P-match, by specifying a code snippet associated with appropriate options. In addition, we can specify the name of an identifier or literal, if we place \$ before the name.

A type 3 code clone  is one with a few statement addition, or deletion, or change for a seed snippet.
Thus the query for type 3 matching could be made from the seed by adding meta-tokens such as \$., \$\$, or \$*, deleting some regular tokens in the seed, or modifying some regular tokens with \$., \$\$, or other meta-tokens, if we could specify how to modify the seed. We call such a  type 3 matching {\em specified type 3 matching}, here.

On the other hand, we might want to make a broad query that  matches all the code snippets of type 3 code clone for the seed with similarity higher than a threshold. We call it {\em unspecified type 3 matching}. Currently, CC matching (and \ccgrep) does not have a feature for the unspecified type 3 matching. We will discuss further on this issue in Sec.\ref{sec:unspecifiedT3}.

Type 4 clones, which are semantically the same but syntactically different, are not considered here, because those cannot be found by syntactical pattern matching approach like ours and we consider that they are out of scope in this research.

\subsubsection{Comparison with \grep}

\vspace{0.2cm}
\noindent
{\bf [Type 1 Matching]}\\
For finding type 1 clones with \ccgrep, we can place a code snippet as the query with non-normalization option (or adding \$ to all identifiers and literals). Note that query strings below are surrounded by a box to clearly separate from other descriptions.
\begin{description}
\item[q21(\ccgrep):]\ \fbox{\tt int a = b;}\ \ with non-normalization option
  \hspace{0.2cm}
\end{description}
Instead, \grep\ needs to care about white spaces between tokens.
\begin{description}
\item[q22(\grep):]\ \fbox{\tt \tb s*int\tb s+a\tb s*=\tb s*b\tb s*;}
\end{description}
As seen here, q21 is  simpler and more straightforward than q22.
Furthermore, additional complication should be needed for q22 if we would want to eliminate comments possibly located between tokens.

\vspace{0.2cm}
\noindent
{\bf [Type 2 Matching]}\\
Following are type 2 matching with any {\tt long} variable declaration.

\begin{description}
\item[q23(\ccgrep):]\ \fbox{\tt \$long a;}
\item[q24(\grep):] \  \fbox{\tt long\tb s+[a-zA-Z][a-zA-Z\_0-9]*\tb s*;}
\end{description}
Both are equivalent but \grep\ requires a much more complex representation for any possible identifier.

\vspace{0.2cm}
\noindent
{\bf [Specified Type 3 Matching]}
\begin{description}
\item[q25(\ccgrep):]\ \fbox{\tt \$time(\$\$)}\ \ Find $time$ with any parameter
  \hspace{0.2cm}
\item[q26(\grep):] \ \fbox{\tt time( }\ \ Find string `$time($'
\end{description}
In this case, we are searching type 3 clones of a function call {\tt time()}, and \ccgrep\ can easily specify it directly. However, in \grep\ case, if we simply give {\tt time}, it matches many variable names and comments, so we add '(', a partial string of function call, which might reduce such unwanted matches.
We have applied these two queries to linux's tools\footnote{\textasciitilde~linux/tools, rev. 4.20.0.}. \ccgrep\ matched 47 {\tt time()} function calls. On the other hand \grep\ matched 382  lines that include {\tt strftime()}, {\tt get\_time()}, and many other function calls\footnote{If we add -w option (word-based matching) to \grep , the result is reduced to 7 matches but it misses the cases immediately followed by a word such as {\tt time(NULL)}.}. Therefore, we would say that our approach is  straightforward to locate specific identifiers or function calls, and that  creating proper queries for  \grep\ is more comprehensive and difficult, compared to our our CC matching approach with \ccgrep.




\subsubsection{Narrowing Matching Results}

An advantage of \ccgrep~ is interactive and repeated matching trials such that the users can try various queries immediately after a result is not sufficient. As an example, we show  in Tab.\ref{tab:kmalloc} a process of investigating a system call function \texttt{kmalloc()} in Linux usb driver sources /drivers/usb/*.

\begin{table}[tb]
    \centering
    \caption{Various Queries for kmalloc in usb drivers}
    \begin{tabular}{|c|| l | r |  } \hline
      &  query  & \#found  \\ \hline \hline
     1  &  \texttt{\footnotesize{kmalloc(\$\$)}} & 109,101  \\ \hline
     2  &  \texttt{\footnotesize{\$kmalloc(\$\$)}} &  333 \\ \hline
     3  &  \texttt{\footnotesize{\$kmalloc(sizeof(\$\$),\$\$)}} & 84 \\ \hline
     4  &  \texttt{\footnotesize{\$kmalloc(sizeof(struct x),\$\$)}} & 29 \\ \hline
     5  &  \texttt{\footnotesize{struct x *p = \$kmalloc(sizeof(struct x),\$\$)}} & 9 \\ \hline
     6\footnotesize{\dag}  & \texttt{\footnotesize{struct u132\_endp *endp =}} &  \\ & \texttt{\footnotesize{~~~~kmalloc(sizeof(struct u132\_endp), mem\_flags);}} & 3 \\ \hline
    \end{tabular} \\
    \footnotesize{\dag With command option for non normalization (type 1 matching)}
    \label{tab:kmalloc}
\end{table}

In this example, we start at query 1 with the type 2 matching of \texttt{kmallock()} with any parameter list denoted by \$\$. This query matches any function calls, thus it produces a large number of the result(\#found). In query 2, we pin down function name to kmalloc, so that the result is reduced to a few hundred. We can further narrow the matches by specifying the first parameter with sizeof at query 3, and with any struct name x at query 4. At query 5, we specify an assignment to a pointer variable with the same struct name x, producing nine matches which are easily checked by hand. We take one of these and give it as query at 6 as it is with non normalization option, resulting in three type 1 clones including the query snippet itself.

As shown by this example, we can interactively and effectively narrow or widen the matching with adding normal or meta-tokens in  queries.

\subsection{RQ2: Accuracy of ccgrep}

For evaluation of  query-matching (or information retrieval) systems, recall and precision values, computed by comparing the matched results with the oracles for the queries, are popularly employed\cite{ModernInfoRetriev}. Here in our approach, however, the query to CC matching has no ambiguity and it  reports the matching result rigorously as expected and specified by the query with options. In such sense, the result is always the same as the oracle, i.e., the recall and precision are always one. Thus, instead of using recall and precision, here we simply investigate if \ccgrep\ works accurately in the sense that code clones already reported by other approaches could be found by our approach.

For this purpose, first we have employed Big\-Clone\-Bench\cite{bigclonebench} that is a huge collection of various kinds of code clones.
We have extracted all pairs classified as type 1 and type 2 code clones from Big\-Clone\-Bench, and for each clone pair $(sp1, sp2)$, we have checked if $sp2$ is successfully found in the result of \ccgrep\ for $sp1$ as query with appropriate options, and vice versa. Tab.\ref{tab:BCBtype12} shows the numbers of type 1 and 2 clones, accurately found and not.

\begin{table}[tb]
    \centering
    \caption{Checked Clones in BigCloneBench}
    \begin{tabular}{|c|| r | r | r| } \hline
    Clone Type  &  Clone Pairs  & Found & Not Found  \\ \hline \hline
    Type 1  &  48116 & 48111 & 5* \\ \hline
    Type 2  &   4234 &  4232 & 2* \\ \hline \hline
    Total   &  52350 & 52343 & 7* \\ \hline
    \end{tabular} \\
    \label{tab:BCBtype12}
    \begin{flushleft}
    \hspace{1.3cm}* indicates faulty clone pairs in BigCloneBench.
    \end{flushleft}
\end{table}

As we can see in Tab.\ref{tab:BCBtype12}, most type 1 and 2 clones are found accurately. There were several cases of not-found clones, and we have investigated further those cases and recognized that those cases are faults of the classification of Big\-Clone\-Bench, some of which should be classified into type 3, and some others are not clones. Thus, we can say that all of proper type 1 and 2 clones in Big\-Clone\-Bench were perfectly found by \ccgrep .

For type 3 clones, since Big\-Clone\-Bench contains a lot of faulty type 3 data, we have instead used CBCD data\cite{LiCBCD2012}, that contains 11 type 3 clone sets taken from the source code of Git, the Linux kernel, and PostgreSQL.
We have crafted type 3 queries from one  of code snippets in each clone set as the seed and have checked if those queries accurately match the other snippets in the same clone set.
We have confirmed that all the crafted queries accurately match other snippets in each clone set. Fig.\ref{fig:type3example} shows an example of a crafted query and its type 3 clone set. In this example, function {\tt map\_write(...)} is repeatedly invoked followed by repeated assignments to variable {\tt chip}'s elements.

\begin{figure}
\footnotesize
\begin{itembox}{ccgrep query}
\begin{verbatim}
$( $map_write($map, $$, $$); $) $+
$( $chip->state = FL_ERASING; $) $+
\end{verbatim}
\end{itembox}
\begin{itembox}{Snippet 1}
\begin{verbatim}
map_write(map, cfi->sector_erase_cmd,
          chip->in_progress_block_addr);
chip->state = FL_ERASING;
chip->oldstate = FL_READY;
\end{verbatim}
\end{itembox}
\begin{itembox}{Snippe 2}
\begin{verbatim}
map_write(map, CMD(0xd0), adr);
map_write(map, CMD(0x70), adr);
chip->state = FL_ERASING;
chip->oldstate = FL_READY;
\end{verbatim}
\end{itembox}
\begin{itembox}{Snippe 3}
\begin{verbatim}
map_write(map, CMD(LPDDR_RESUME),
          map->pfow_base + PFOW_COMMAND_CODE);
map_write(map, CMD(LPDDR_START_EXECUTION),
          map->pfow_base + PFOW_COMMAND_EXECUTE);
chip->state = FL_ERASING;
chip->oldstate = FL_READY;
\end{verbatim}
\end{itembox}
\caption{An Example of Crafted Query and Type 3 Clones for CBCD Data}
    \label{fig:type3example}
\end{figure}

As far as our investigation, all the matches are controlled by the query, and are performed accurately as we have expected.

\subsection{RQ3: Performance of ccgrep}
\label{sec:evalperformance}

It is interesting to know that our approach, i.e., token-based and naive sequential pattern matching, can be implemented fast enough for practical use.
We have examined various queries for \ccgrep\ with the target source files of Antlr and Ant in Java, and CBCD data (Git, PostgreSQL, and Linux Kernel)\footnote{For comparison to CBCD, we have used the same data set, but currently those projects are enhanced  2 to 4 times larger, and the results for applying to the current projects grow almost linearly in both \#found and time.} in C, and have measured the performance of \ccgrep .
Following are employed queries. All execution was made with default setting of \ccgrep~ except for the language option.

\begin{description}
\item[qA:] \fbox{\tt a < b? a: b}  \\Find ternary operation to give a smaller value.

\item[qB:] \fbox{\tt T1 f(T2 a) \{ return \$\$; \}}
  \\Find function definition immediately returning a value.
\item[qC:] \fbox{\tt f(\$\$, \$\$, \$\$); }
  \\Find three parameter function.
\item[qD:] \fbox{\parbox[bt]{5cm}
                {\tt for(a = 0; a < \$\$; a++) \{ \$\$ \} \\
                 \$| \\
                a = 0; while(a < \$\$) \{ \$\$ a++; \}}}
                \\Find {\tt for} or {\tt while} statement with a control variable.

\end{description}

Tab.\ref{tab:TargetAndResult} shows the size metrics of the target, the number of found snippets, and the execution time of each query on a workstation with Intel Xeon E5-1603v4 (@2.8GHz $\times$ 4), 32GB RAM, and Windows 10 Pro for WS 64bit.

As we can see from Tab.\ref{tab:TargetAndResult}, the execution times are about 10 sec. even for a few million lines of  Linux kernel target. This might not be very fast, but we would think that they are acceptable speed as a daily-used development or maintenance tool.

The execution times for qA to qD are very stable for each target. For example, in the case of Linux, they are about 10 sec. even for the small \#found case (48 for qA) and the large \#found case (187,653 for qC). Thus, we would say that the execution time is not heavily affected by the result size (\#found) but mainly affected by the target size (\#line).

Targets Ant in Java and PgSQL in C have similar sizes around 140-180K lines, and the execution times are also similar around 1-1.5 sec. This would show that the execution time is not strongly affected by the target language.



\begin{table}[tb]
    \centering
    \caption{Target and Execution Result by \ccgrep }
    \smaller
    \begin{tabular}{|l|l||r|r|r|r|r|} \hline
        \multicolumn{2}{|l||}{Target} & Antlr & Ant & Git & PgSQL & Linux\\ \hline  \hline
        \multicolumn{2}{|l||}{Lang.} & Java & Java & C & C & C \\  \hline
        \multicolumn{2}{|l||}{\#file} & 678 & 1,272 & 339 & 904 & 15,123 \\  \hline
        \multicolumn{2}{|l||}{\#line}  & 59,511 &    138,396  & 90,495 & 177,174 & 3,756,212 \\ \hline \hline
        qA & \#found    & 0     & 2     & 8     & 3      & 48 \\ 
               & time(sec.) & 1.12  & 1.32  & 1.11  & 1.43   & 9.46 \\ \hline
        qB & \#found    & 159   & 161   & 7     & 27     & 543 \\ 
               & time(sec.) & 1.15  & 1.33  & 1.10  & 1.47   & 10.15 \\ \hline
        qC & \#found    & 1,710 & 2,487 & 5,717 & 10,603 & 187,653 \\ 
               & time(sec.) & 1.20  & 1.38  & 1.13  & 1.55   & 12.01 \\ \hline
        qD & \#found    & 1     & 13    & 442   & 621    & 10,754 \\ 
               & time(sec.) & 1.19  & 1.52  & 1.10  & 1.49   & 11.06 \\ \hline
      \end{tabular} \\
    \flushright{Antlr: Antlr4 v.4.7.2, Ant: Apache Ant v.1.10.5, Git: v.1.6.4.3,\\
    PgSQL: PostgreSQL v.6.5.3, Linux: Linux kernel v.2.6.14rc2}
    \normalsize

      \label{tab:TargetAndResult}
\end{table}


For comparison to other tools, we have used  a code snippet finder NCDSearch\cite{IshioMSI18} and  \grep.
NCDSearch finds similar code blocks in the target file for the query code block, by checking the normalized compressed distance of the query and target blocks, and it reports sometimes false positive results. For NCDSearch, we have used qA as its query, and for \grep\, we have used GNU grep\cite{grepGnu} with following qA'.

\begin{description}
\item[qA'({\tt grep}):] \fbox{\parbox[bt]{5.6cm}
                {\tt ([a-zA-Z\_][a-zA-Z\_0-9]*)\tb s*< \\
                ([a-zA-Z\_][a-zA-Z\_0-9]*)\tb s*\tb ?\tb s* \\
                \tb 1\tb s*:\tb s*\tb 2}}

\end{description}
This qA' is to find ternary expression using an extended regular expression\footnote{We have used options {\tt -w} and {\tt --include='*.[ch]'}. Some version of \grep\ might require -E option or {\tt egrep} command for the extended regular expression.}, and it is almost equivalent query to qA for \ccgrep\, except that qA' does not allow the new line between tokens. In order to allow the new lines for \grep , the query expression becomes too complex to present here.

Tab.\ref{tab:toolcomparison} shows the execution result of \ccgrep, NCDSerach, and  \grep.
Since NCDSearch contains false positives, the number of found (\#found) is  larger than that of \ccgrep\ that contains no false positives. Also, since qA' could sometimes miss the snippets with new lines, \#found for \grep\ is sometimes smaller than that of \ccgrep.

\begin{table}[tbp]
    \centering
    \caption{Comparison of \ccgrep\ with NCDSearch and  \grep\ }
    \smaller
    \begin{tabular}{|l|l||r|r|r|r|r|} \hline
        \multicolumn{2}{|l||}{Target} & Antlr & Ant & Git & PgSQL & Linux\\ \hline \hline
        qA       & \#found    & 0    & 2     & 8    & 3     & 48 \\
        ~(ccgrep)    & time(sec.) & 1.12 & 1.32  & 1.11 & 1.43  & 9.46 \\
                     & time ratio      & 1.0  & 1.0   & 1.0  & 1.0   & 1.0 \\ \hline
        qA       & \#found    & 3    & 21    & 22   & 80    & 21,047 \\
        ~(NCDSearch) & time(sec.) & 5.55 & 12.00 & 8.41 & 16.54 & 366.39 \\
                     & time ratio      & 4.96 & 9.09  & 7.58 & 11.57 & 38.73 \\ \hline
        qA'      & \#found    & 0    & 1     & 8    & 2     & 44 \\ 
        ~(grep)      & time(sec.) & 0.24 & 0.30  & 0.12 & 0.24  & 2.62 \\ 
                     & time ratio      & 0.21 & 0.23  & 0.11 & 0.17  & 0.28 \\ \hline
    \end{tabular}\\
    \flushright{Time ratio means the ratio of the execution times of each tool to \ccgrep.}
    \normalsize
    \label{tab:toolcomparison}
\end{table}

\ccgrep\ is faster than NCDSearch for all targets,  and \grep\ is  faster than \ccgrep . Since \grep\ is known to be very fast\footnote{https://lists.freebsd.org/pipermail/freebsd-current/2010-August/019310.html},  we would think that the speed of \ccgrep\ is  acceptable  as a practically usable software engineering tool. We will also discuss on further improvement of the performance in Sec.\ref{sec:DiscussSpeed}.

\section{Discussions}

\subsection{CC Matching Approach}

We have proposed and formulated CC matching as a method of finding code snippets for the user's interest. The approach is based on the notion of finding code clones in the target, and the query is a simple code snippet or its extension with meta symbols representing specified or wildcard tokens, regular expressions, and so on.

We would think that our approach is a very good support for software maintainers who need to look around huge source code, due to various reasons such as bug fixing, feature locating,  refactoring and so on. Compared to \grep\, CC matching (and \ccgrep\ ) generally can make rich queries more easily and compactly, in the sense that white spaces, new lines, or comments are not considered, and matching identifier or literals can be controlled with meta-tokens.

In fact, during our testing and evaluation of \ccgrep\ we scanned the source code of Linux for specific clones. We
discovered that the pattern \mycode{a<b?a:b} had been removed over time, yet some instances persisted. As
described in the Motivating Example, we fixed 3 of them in the \mycode{usb/drivers} modules and submitted patches for them. To this
date, two of the three have been accepted and are already integrated into Linux.

There are many other approaches proposed to make rich queries for code matching\cite{AbouAssaleh2004SurveyOG}; however, those are not easy to use for many software engineers due to their own query forms. Our approach relies on the notion of finding code clones, which is straightforward and  easy to understand and to use for many people.
In addition, we have adopted  a \grep-like interface for \ccgrep, which would greatly reduce the burden of using a new tool.

Issues of code clones have become popular and acknowledged by not only by software engineering researchers but also
industry people\cite{koschke_et_al:DR:2012:3477}, along with prevalence of various clone detectors such as stand-alone
tools, CCfinderX\cite{ccfinderx}, NiCad\cite{Cordy2011NiCad}, and SourcererCC\cite{Sajnani2016SourcererCC}, or  IDE's
with clone detection features such as Eclipse\cite{DBLP:conf/sac/ZibranR12}, or Visual
Studio\cite{visualstudio}. However, those are basically large systems, and their proper installation and operation are
not easy in general. We would encourage the creators of these tools to create simpler interfaces that make it easy to
find specific instances of clones of small snippets, as ccgrep does.

\subsection{Extending Matching for Unspecified Type 3 Matching}
\label{sec:unspecifiedT3}

As discussed in former sections, CC matching allows type 3 queries as the specified type 3 matching, where we have to predict and specify the variable parts of the seed snippet of the clone pair, with some wildcard tokens. For the unspecified type 3 matching, i.e., if we do not know the variable parts clearly, but would want to find just 'similar' code snippets with a similarity metric value higher than a threshold, the current CC matching approach might be insufficient.

To solve this issue, we could  extend CC matching to allow unspecified type 3 queries, by introducing, say, similarity-based matching or error-allowable matching\cite{Navarro2001}. However, this introduction would be far away from the current policy of CC matching such that the matches are rigorously controlled by the query with options and no ambiguity in the result is allowed.
In such sense, although we are interested in this extension, it would be accomplished as  another matching framework and a different tool.

\subsection{Performance of ccgrep }
\label{sec:DiscussSpeed}

As shown in the evaluation, \ccgrep\ is not as fast as \grep~ but we think that it is acceptable as a practical and useful tool for finding code snippets.
Currently, \ccgrep\ employs a simple and naive sequential matching algorithm. The mismatch information is not used for the following process, but the algorithms using the mismatch information such as  Knuth-Morris-Pratt algorithm or Boyer-Moore algorithm\cite{ComputingPatterns} could be used for further performance improvement.
In addition, in the current implementation, a sequential trial process for the selection of regular expressions is employed, but it could be improved by introducing the parallel processes.


\section{Related Works}

\subsection{Pattern Matching Tools}

\grep\ was originally developed as a simple  pattern matching tool for Unix, and has been enhanced with many regular-expression features and other fast pattern matching algorithms for GNU grep. Variants of grep, such as context grep cgrep, approximate grep agrep, and many others had been proposed and implemented to meet various requirements\cite{AbouAssaleh2004SurveyOG}. However, there is no one for clone-based matching like ours.

Semantic-based matching tool sgrep\cite{sgrep2002} relates to our work in the sense of matching based on the program contexts, and the logic-based query pattern capturing language is proposed in \cite{sivaraman2019active}. However, compared to these approaches, our approach is much closer to the original program syntax and code snippet, than their proprietary query patterns. We would think, the learning cost for our approach is smaller than those for special and proprietary matching patterns.

Formalization of abstracted pattern matching over various languages had been proposed by Dekel et al.\cite{Dekel2003-pattern-lang}. They have defined an abstract code pattern language CPL focusing on semantics rather than syntax.
Paul et al. had proposed  a formal data model with an algebraic expression-based query language\cite{Paul-1994-query}.
Those approaches are aiming at generalization and formalization of query patterns, and our approach, on the other hand, focuses on  language-depend  easily-created query patterns with notion of code clone and meta-pattern.

awk is a pattern matching and text processing tool for general text handling\cite{awk}. Though it provides flexible pattern matching with regular expressions and powerful actions associated with the matches, no mechanism for matching based on the notion of code clones is  provided.

\subsection{Code Clone Detectors}

There are numerous number of publications on code clone detection methods and their tools\cite{rattan2013software,ROY2009470}. Token-based approach is a very popular process for clone detection, and  one of early works was   Dup\cite{Baker92aprogram}, where important notions such as P-match and suffix tree matching algorithm were used. CCFinder extended this idea to strengthen practical use by normalization and other transformation\cite{Kamiya2002CCFinder}. CP-Miner used a frequent subsequent  mining method to find similar sequences\cite{CPminerli2006cp}.

Textual-based approach is another major method, pioneered by Johnson\cite{Johnson1993}, and it has been used by many others\cite{ducasse1999language}. NiCad used both text and tree-based approach to detect near-miss clones\cite{Cordy2011NiCad}.
SourcererCC employed block-based matching and heuristics for filtering out redundant comparison to scale analysis\cite{Sajnani2016SourcererCC}.

Most of these approaches focus on finding all code clone pairs in the target file collection. They report all code clones or similar code snippets with similarity higher than certain threshold. Precisely controlling the matches with meta-symbols like CC Matching cannot be accomplished by those approaches.


\subsection{Code Snippet Search Tools}

There are several tools specialized for finding code snippet.
CBCD has been designed for finding related code snippets from a buggy code snippet, by using  matching of Program Dependence Graph (PDG)\cite{LiCBCD2012}. It can be used to find type 1, 2, and  3 clones; however, the matching generally requires long pre-processing time to construct PDG, and so this approach would not fit to the handy clone finding that we are interested in. For example, it is reported that pre-processing time for a part of the Linux kernel with 170K LOC required 32 minutes, which is far slower than our approach, even though it is claimed that the pre-processing is a one-time overhead and repeatedly used for many queries.

NCDSearch has been designed to find similar code snippets in the pile of source code files for the analysis of code reuse and evolution\cite{IshioMSI18}. The query and the target snippets in block level are compressed together to see the similarity. The approach would be unique and interesting, but the speed is slower than ours as shown in Sec.\ref{sec:evalperformance}. In addition, the user cannot predict properly the similarity to be matched by the approach, which produces false-positive and unexpected matching result.

Siamese has been developed for finding code clone pairs for a query method or file using  multiple representation of n-gram token sequences with inverted index\cite{Chaiyong2019Siamese}. It reports the ranked result of type 1, 2, and 3 clones, and it shows  good accuracy and scalability. However, although its query response time is small, it requires fairly long indexing time (e.g., about 10 minutes indexing time for 10,000 method target). Thus its application and usage would be different from ours.

\section{Threat to Validity}

We have started with the issue on the current pattern matching tools and code clone tools. For highly experienced users of those tools, our issue might not be applicable due to their deep knowledge and skill of the tools. We are interested in an approach widely applicable to many software engineers including not highly experienced ones for those tools. Controlled experiment and/or feedback from various users might help to understand the issue deeply.

Our evaluation in this paper was mainly focusing on finding code snippets with well expected and specifiable patterns, as discussed in Sec.\ref{sec:unspecifiedT3}.
Experiment with broader or vaguer queries including unspecified type 3 matching might lower performance, but we would think that handling such queries will be a different research topics including a different tool implementation.

Expressiveness for the specified type 3 matching might be non-exhaustive. However, as long as we have investigated the CBCD clone data, the queries for all type 3 clones were created without any difficulty and we strongly think that we can easily extend them to many other type 3 clones.

\section{Conclusions}
We have presented and formulated a new approach to find code snippets in the target files with notion of code clone and meta-pattern, as CC matching, associated with its implementation \ccgrep. It is a practical and effective pattern matching method,  and the tool is efficient and easy-to-use to many software engineers.

As a future direction, we are interested in  further performance improvement by using more efficient pattern matching algorithms as discussed in Sec.\ref{sec:DiscussSpeed}.  Also, we are trying to spread use of \ccgrep~ to industry and academia. We believe that \ccgrep~ is a very good tool to explore similar bugs to prevent the late propagation\cite{barbour2011ICSM}, and some companies have already shown their interest to their industry applications.

\newpage

\bibliographystyle{ACM-Reference-Format}
\bibliography{ccgrep-bib}

%

\end{document}

%% file: macros.tex
\usepackage{xspace}
\usepackage{xcolor}

\ifdraft
    \usepackage[colorinlistoftodos]{todonotes}
    \newcommand{\dmg}[1]{{\color{blue}\emph{dmg Says: #1}}\xspace}
    \newcommand{\dmgtodo}[1]{{\color{blue}\emph{dmg Todo: #1}}\xspace}
    \newcommand{\inoue}[1]{{\color{cyan}\emph{inoue Says: #1}}\xspace}
    \newcommand{\inouetodo}[1]{{\color{cyan}\emph{inoue Todo: #1}}\xspace}
\else
    \usepackage[disable]{todonotes}
    \newcommand{\dmg}[1]{}
    \newcommand{\dmgtodo}[1]{}
    \newcommand{\inoue}[1]{}
    \newcommand{\inouetodo}[1]{}
\fi

\newcommand{\mycode}[1]{\texttt{\small #1\xspace}}
